\documentclass[%
 superscriptaddress,
preprint,
preprintnumbers,
nofootinbib,
 amsmath,amssymb,
 aps,
floatfix,
]{revtex4-2}

\usepackage[english]{babel}
\usepackage{graphicx}
\usepackage{dcolumn}
\usepackage{bm}
\usepackage{siunitx}
\usepackage[version=4]{mhchem}
\usepackage{amsmath}
\usepackage{lineno}
\usepackage{url}

\begin{document}

\title{Emergent Chaos-Like Dynamics of Spin--Orbit-Torque-Driven Magnetic Transitions}

\author{L.-M.\ Kern*$^\dagger$}
\affiliation{Max Born Institute for Nonlinear Optics and Short Pulse Spectroscopy, 12489 Berlin, Germany}
\email[Corresponding author: ]{kern@mbi-berlin.de}
\author{K.\ Litzius$^\dagger$}
\affiliation{Experimental Physics V, Center for Electronic Correlations and Magnetism, University of Augsburg, 86159 Augsburg, Germany}
\thanks{These authors contributed equally to this work.}
\author{V.\ Deinhart}
\affiliation{Helmholtz-Zentrum Berlin für Materialien und Energie, 14109 Berlin, Germany}
\affiliation{Ferdinand-Braun-Institut (FBH), 12489 Berlin, Germany}
\author{M.\ Schneider}
\affiliation{Max Born Institute for Nonlinear Optics and Short Pulse Spectroscopy, 12489 Berlin, Germany}
\author{C.\ Klose}
\affiliation{Max Born Institute for Nonlinear Optics and Short Pulse Spectroscopy, 12489 Berlin, Germany}
\author{K.\ Gerlinger}
\affiliation{Max Born Institute for Nonlinear Optics and Short Pulse Spectroscopy, 12489 Berlin, Germany}
\author{R.\ Battistelli}
\affiliation{Helmholtz-Zentrum Berlin für Materialien und Energie, 14109 Berlin, Germany}
\author{D.\ Metternich}
\affiliation{Helmholtz-Zentrum Berlin für Materialien und Energie, 14109 Berlin, Germany}
\author{D.\ Engel}
\affiliation{Max Born Institute for Nonlinear Optics and Short Pulse Spectroscopy, 12489 Berlin, Germany}
\author{C.\ M.\ Günther}
\affiliation{Technische Universität Berlin, Zentraleinrichtung Elektronenmikroskopie, 10623 Berlin, Germany}
\author{M.-J.\ Huang}
\affiliation{Deutsches Elektronen-Synchrotron, 22607 Hamburg, Germany}
\author{K.\ Höflich}
\affiliation{Ferdinand-Braun-Institut (FBH), 12489 Berlin, Germany}
\affiliation{Helmholtz-Zentrum Berlin für Materialien und Energie, 14109 Berlin, Germany}
\author{F.\ Büttner}
\affiliation{Experimental Physics V, Center for Electronic Correlations and Magnetism, University of Augsburg, 86159 Augsburg, Germany}
\affiliation{Helmholtz-Zentrum Berlin für Materialien und Energie, 14109 Berlin, Germany}
\author{S.\ Eisebitt}
\affiliation{Max Born Institute for Nonlinear Optics and Short Pulse Spectroscopy, 12489 Berlin, Germany}
\affiliation{Technische Universität Berlin, Institut für Optik und Atomare Physik, 10623 Berlin, Germany}
\author{B.\ Pfau}
\affiliation{Max Born Institute for Nonlinear Optics and Short Pulse Spectroscopy, 12489 Berlin, Germany}

\begin{abstract}
\textbf{Spin--orbit torques (SOTs) are widely used to control magnetization in nanoscale electric systems and are typically assumed to drive skyrmion nucleation and motion in a deterministic manner, especially in materials with strong Dzyaloshinskii--Moriya interaction. Here, using time-resolved holography-based x-ray microscopy supported by micromagnetic simulations, we reveal that on nano- to picosecond timescales the actual dynamics can deviate strikingly from this expectation by producing transient regimes of chaos-like behavior. By exploiting deterministic skyrmion generation at an anisotropy-engineered defect and implementing a high-resolution pump--probe scheme, we directly track the magnetization evolution in real space. This approach uncovers a dynamic phase transition that separates coherent SOT-driven motion from a regime of transient instability characterized by picosecond-scale fluctuations, strong domain disorder, topological instabilities, and skyrmion shedding, experimentally observed here for the first time. During SOT actuation, the system briefly enters this instability regime, showing short-lived chaos-like behavior, yet it reliably relaxes into robust and reproducible final states. Our results demonstrate a powerful methodology for accessing time-averaged nano- to picosecond dynamics in magnetic systems and reveal a previously hidden layer of transient, topologically rich behavior underlying nominally deterministic skyrmion control.}
\end{abstract}

\maketitle

\section*{Introduction}
Spin--orbit torques (SOTs) are among the most effective mechanisms for manipulating magnetization via electric current~\cite{hellman2017interface,shao2021roadmap}. These torques enable versatile control over magnetization in nanoscale electric systems, powering applications from magnetic tunnel junction switching~\cite{baumgartner2017spatially,krizakova2022spin}, domain-wall motion~\cite{caretta2018fast,avci2019interface}, and spin-wave generation~\cite{fulara2019spin, demidov2020spin}, to the creation and manipulation of topological textures like magnetic skyrmions~\cite{woo2016observation,buttner2017field,legrand2017room}. SOTs are especially valued for their ability to drive reliable, fast, and well-controlled dynamics, a crucial requirement for spintronic technologies~\cite{yang2021chiral}. These properties, supported by micromagnetic simulations, have positioned SOTs as ideal engines for steady-state motion in racetrack devices based on materials with strong Dzyaloshinskii--Moriya interaction (DMI)~\cite{fert2013skyrmions,litzius2017skyrmion,litzius2020role,zhang2025nanofluidic}. However, probing the sub-\SI{100}{nm}, picosecond-scale response of magnetization under SOT remains a formidable challenge, previously limited to confined geometries \cite{buttner2015dynamics,jiang2015blowing}, localized contacts~\cite{finizio2019deterministic}, or isolated natural defects~\cite{woo2017spin,woo2018deterministic}. 

Recent advances in local anisotropy modification using focused ion beam irradiation now allow precise skyrmion generation and motion in racetrack geometries~\cite{juge2021helium,kern2022deterministic}. Over the past years, the role of pinning has shifted from a detrimental effect~\cite{reichhardt2022statics,ohara2021confinement} to a deliberately engineered mechanism promoting chirality exchange~\cite{zhang2023chiral}, diffusion~\cite{zhang2025asymmetric}, and transformation of topologically non-trivial textures~\cite{xia2024transformation} for future spintronics applications. 

Here, we combine such engineered nanometer-sized nucleation sites inside a racetrack with time-resolved x-ray holography to directly visualize magnetization changes with nanometer and picosecond resolution during and after a nanosecond current-induced spin--orbit torque pulse. Our observations uncover a previously inaccessible regime: under current-induced SOT action, we witness a cascade of rapidly evolving, spatially disordered fluctuations, signaling a dynamic phase transition (DPT)~\cite{riego2018towards}. A magnetic DPT occurs when varying a system parameter---here, the spin--orbit torque drive---induces a qualitative change of the magnetization dynamics. This transition is characterized by a threshold separating a stable, coherent response from a regime dominated by topological reconfigurations and extreme sensitivity to initial conditions, suggesting transient chaos. The presence of such a transition marks the breakdown of conventional predictability, even as the system converges on repeatable final states. Among the rich dynamics we observe is skyrmion shedding, long predicted under spin-transfer torque~\cite{everschor2017skyrmion} but imaged here for the first time. Our results highlight a deeper complexity in SOT-driven systems than previously appreciated, in which instability promotes the emergence of transient, complex spin configurations, potentially serving as an effective mechanism to control switching. Our findings lay the groundwork for leveraging dynamic instabilities and critical behavior in future spintronics technologies.

\section*{Results}

\begin{figure*}
    \begin{center}
		\includegraphics[width=1.0\linewidth]{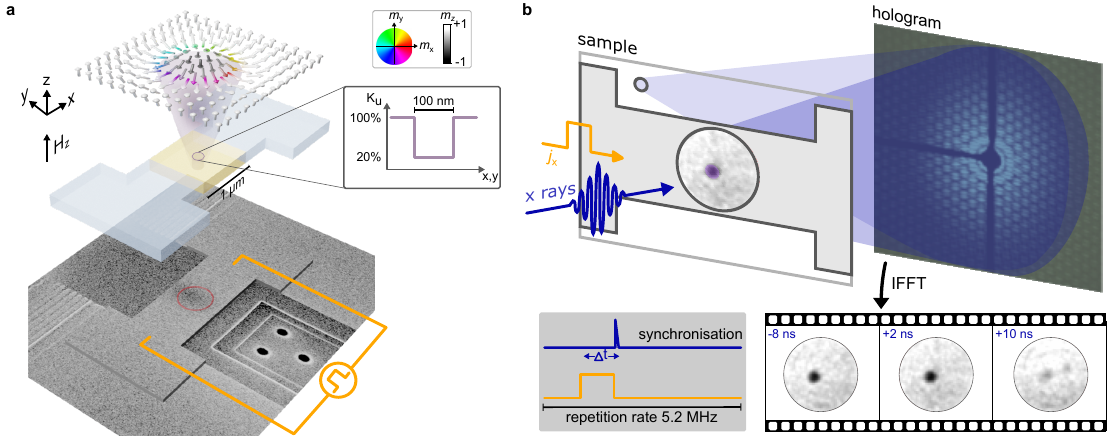}
	\end{center}
	\caption{\label{fig:setup}\textbf{Observation of SOT-driven magnetization dynamics in simulation and experiment.} \textbf{a} \textit{Bottom}: Scanning electron microscopy (SEM) image of the magnetic racetrack device with the holographic field of view (red circle) and the electric connection for pulse injection indicated. \textit{Center}: Model of the device used for the simulation with the \SI{1}{\micro m}$\times$\SI{1}{\micro m} large area containing the anisotropy-engineered region in the centre. \textit{Top}: Illustration of the spin configuration in the dot after nucleating a skyrmion. \textbf{b} Schematic view of the pump-probe scheme for x-ray holography imaging of the current-induced SOT-driven magnetization dynamics at an ion-irradiated region.}
\end{figure*}

We study, both experimentally and via simulation, an asymmetrically stacked ferromagnetic multilayer comprising \num{15} repeats of a Pt/Co$_{60}$Fe$_{25}$B$_{15}$/MgO trilayer with perpendicular magnetic anisotropy (PMA) and interfacial DMI (see Methods). This standard system is widely used to investigate SOT-driven skyrmion creation and motion~\cite{buttner2017field,gerlinger2021application,litzius2017skyrmion,lemesh2018current,litzius2020role}. In our experiment, the magnetic material was structured into a racetrack geometry to allow for current-pulse injection~\cite{buttner2017field,kern2022deterministic}, as shown in Fig.~\ref{fig:setup}. The key feature of the device is a \SI{100}{nm} wide He$^+$-ion irradiated circular region near the center of the racetrack, which exhibits reduced PMA and locally enhanced SOT efficiency~\cite{yun2019lowering,dunne2020helium}, thus allowing us to address this region selectively with SOT~\cite{kern2022deterministic}. Furthermore, the anisotropy-engineered region supports stable skyrmions at magnetic fields where the pristine film remains in saturation, consistent with earlier findings~\cite{kern2022deterministic}. This anisotropy-engineered region (from now on referred to as the ``dot'') hence allows us to control the SOT-driven dynamics spatially and to ensure a deterministic skyrmion generation.

\subsection{Modelling SOT-driven dynamics of local magnetization}

\begin{figure*}
	\includegraphics[width=1.0\linewidth]{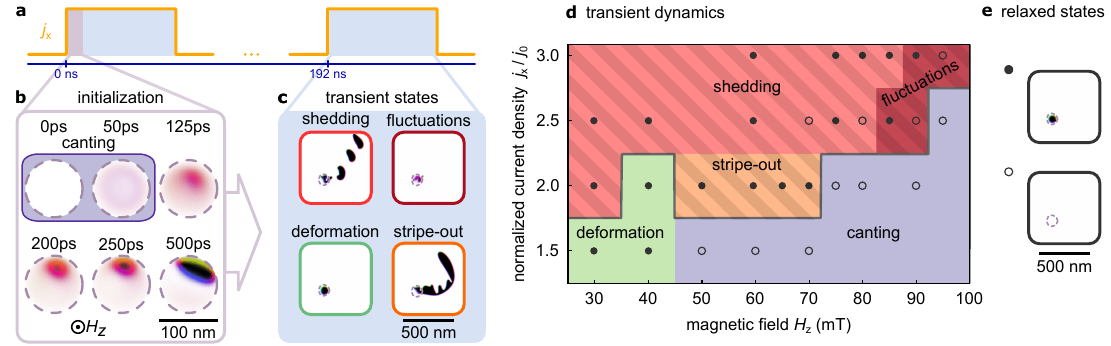}
	\caption{\label{fig:phase_diagram}\textbf{Micromagnetic modelling reveals prototypical types of magnetization dynamics.} \textbf{a} Two \num{10}-ns current pulses are schematically shown. \textbf{b} The  initialization pulse (potentially) switches the magnetization in the dot, showing an extended switched bubble at \SI{500}{ps}. \textbf{c} The second pulse mimicks a cycle in the pump-probe scheme, revealing five different types of dynamics as a function of the magnetic field and the current density: canting, deformation, stripe-out, shedding and internal fluctuations. All images are cropped to the same region of interest, with a dashed outline indicating the dot. \textbf{d} Phase diagram of corresponding dynamics in the parameter space of applied magnetic field and current density. Markers represent parameter values for which simulations were carried out. Magnetization fluctuations are observed in the stripe-shaded region. See Methods for details on $j_0$. \textbf{e} The relaxed state is either (i) a bubble nucleated in the dot (filled markers in d), or (ii) a homogeneously magnetized state (open markers).}
\end{figure*}

To realize the pump--probe experiment schematically illustrated in Fig.~\ref{fig:setup}, the magnetization dynamics must be repeatable, meaning that the system must at least return to the same initial state before each SOT pump pulse. To identify suitable experimental conditions, we performed micromagnetic simulations of the dynamics initiated at a nanometer-scale dot with modified magnetic properties. The simulated area encompasses the full width and central portion of the racetrack, as shown in Fig.~\ref{fig:setup}a. All simulations were performed at zero temperature, with material parameters kept constant throughout (see Methods for further details).

The simulation involved two successive current pulses of \SI{10}{ns} duration inducing SOT (Fig.~\ref{fig:phase_diagram}a). The first pulse primes the system---typically through nucleation---into a reproducible initial state. This initial state will only be accessible in simulations, as the pump--probe scheme provides access only to time-averaged features in the experiment. During the first \SI{0.5}{ns} (the initialization phase, akin to the first pulse in a long pump--probe sequence), dynamics remain confined to the dot, and we present snapshots cropped to the size of the dot in Fig.~\ref{fig:phase_diagram}b. The dynamics always starts with a coherent spin canting in the whole dot. Above a nucleation threshold, a spin wave emerges from the border of the dot, almost symmetrically propagates towards the center of the dot, and, finally, leads to a completely reversed area at the upper right edge of the dot. This reversed area finally relaxes to a magnetic bubble with the skyrmion number $Q=0$ or $Q=1$ (skyrmion) in the dot. The initial spin canting is generic and does not depend on the spin configuration emerging subsequently. Below the nucleation threshold, the magnetization does not switch, and the dot quickly returns to uniform magnetization.  

The second pulse triggers dynamics from this initial state; following the subsequent evolution thus simulates one typical pump--probe cycle. In Fig.~\ref{fig:phase_diagram}c, we map out different types of dynamics starting from an initial state and present a characteristic frame for each type. Full movies are provided as Supplementary Material. We explore the influence of the current density, $j_x$, and the static applied field, $\mu_0H_z$, which we use to control the dynamics in the experiment, in Fig.~\ref{fig:phase_diagram}d.  We first summarize the results from the simulation before presenting experimental and simulation results side by side in the next section: If the initial state is uniformly magnetized, the second pulse again triggers the spin canting, persisting during the entire pulse duration and then quickly returning to a homogeneous magnetization. If a magnetic bubble was nucleated during the initialization, the bubble experiences only gentle perturbation from the current-induced SOT pulse at low $j_x$ and low $\mu_0H_z$. This perturbation leads to breathing and mild compression mainly along the direction of $j_x$. We denote this dynamics as the deformation dynamics. In contrast to this regime, the dynamics becomes much more violent inside the dot when increasing the current density. The response over a large region in our diagram involves strong magnetic fluctuations and instabilities on picosecond timescales. At moderate applied fields, we additionally observe magnetic fluctuations extending outside the dot, which are highly mobile in the material. Complex, periodically recurring states and transient instabilities of the magnetization characterize these dynamics. They at first manifest as elongations originating from the dot (\textit{stripe-out}), which at even higher current density break off and form small, topological textures that move away from the dot. These textures propagate under an average angle of \num{47}°. Similar magnetization dynamics were already found in models with spin-transfer torque, referred to as skyrmion \textit{shedding}, first introduced by Everschor-Sitte et al., for which a current-driven instability at a magnetic inhomogeneity periodically nucleates and expels topological spin textures \cite{everschor2017skyrmion}. This process is central to skyrmion dynamics at nano- to picosecond timescales, as it represents a repeatable mechanism for generating and propagating topological solitons under strong spin torque drives. In our system, an analogous mechanism is realized via spin--orbit torque, which provides higher efficiency than spin-transfer torque, while experimental constraints require pulsed (pump--probe) excitation rather than continuous direct current operation. Given the close correspondence of a detaching texture from a defect, we therefore identify our observations as skyrmion shedding.

Eventually, the magnetization relaxes into either a homogeneous magnetization (non-switched state) or a bubble (switched) state (Fig.~\ref{fig:phase_diagram}e). Crucially, these states are reliably restored across a broad range of parameters, with only a few exceptions. These exceptions occur in particular at the boundaries of the phase diagram. The reproducibility of the initial and final states predicted in the simulation suggests that meaningful pump--probe imaging can be performed, which is discussed in comparison to the simulation results in the following.

\subsection{Time-resolved experimental observation of SOT-induced dynamics}

\begin{figure*}
	\includegraphics[width=0.48\linewidth]{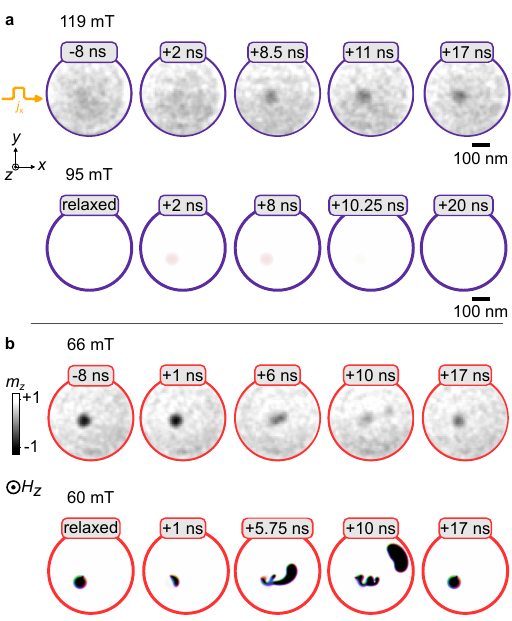}
	\caption{\label{fig:exp+sim}\textbf{Time-resolved imaging of magnetization dynamics.} \textbf{a} Coherent spin canting inside the dot does not switch the magnetization. \textbf{b} A detachment of transient topological textures from the switched dot marks the experimental evidence for skyrmion shedding. The applied current density used in the pump-probe experiment (top row in each panel) and simulation (bottom row) are $j_x=j_0=\SI{4.3e11}{A \per m^2}$ and $j_x=2.5j_0$, respectively. The colour code corresponds to the dynamics introduced in Fig.~\ref{fig:phase_diagram}. The field of view in the experiment is fixed to a circular region (diameter \SI{600}{nm}) (see Methods), and simulation results were cropped to the same size. The entire movies are available in the Supplementary Material.}
\end{figure*}

We directly imaged SOT-driven dynamics using time-resolved x-ray holography combined with numerical phase retrieval at sub-\SI{20}{nm} spatial resolution~\cite{battistelli2023coherent} (see Methods), significantly surpassing previous time-resolved imaging approaches~\cite{buttner2015dynamics,woo2017spin,litzius2017skyrmion,woo2018deterministic,finizio2019deterministic}. We recorded images at various delays $\Delta t$ between the onset of the \SI{10}{ns} rectangular current pulse driving the dynamics and the 100-ps x-ray pulse probing the transient magnetic configuration, capturing frames before, during, and after the current pulse reached the racetrack. Each image reflects the real-space and real-time magnetic state at a specific delay, averaged over \num{4.7e7} pump--probe cycles (see Methods), during which the applied field remains static. To avoid cumulative effects, images were intentionally acquired in a non-sequential order with respect to the delay steps, and static reference images of the unpumped domain configuration were regularly recorded to confirm that the sample remained undamaged throughout the experiment.

As shown in Supplementary Fig.~1c, we apply unipolar rectangular current pulses of \SI{10}{ns} duration, which are known to nucleate skyrmions in comparable multilayer stacks~\cite{buttner2017field,woo2018deterministic,kern2022deterministic}. Before each time-resolved measurement, the sample is saturated at $\mu_0 H_{\mathrm{z}} = \SI{300}{mT}$ to erase magnetic history. To avoid cumulative effects, images were intentionally acquired in a non-sequential order with respect to the delay steps, and static reference images of the unpumped domain configuration were regularly recorded to confirm that the sample remained undamaged throughout the experiment.

By tuning $\mu_0H_z$ and $j_x$, we explored the map of SOT-induced dynamics in our device and found experimental evidence for the different prototypical types of dynamics predicted in our simulations, with two scenarios presented in Fig.~\ref{fig:exp+sim} and more examples in the Supplementary Fig.~2. The static out-of-plane field $\mu_0 H_\mathrm{z}$ and the applied current density for each dataset are indicated in the respective figure captions. Each example includes five representative experimental snapshots alongside corresponding simulation frames that qualitatively match our experimental results for equivalent applied fields and current densities (see Methods for details on comparing experimental and simulation parameters). Full movies are provided in the Supplementary Material.

The first time series in Fig.~\ref{fig:exp+sim}a was recorded near saturation field such that skyrmion nucleation in the dot is suppressed~\cite{kern2022deterministic}. Still, we observe transient changes of the out-of-plane magnetization inside the dot emerging during and lasting for up to twenty nanoseconds after applying the current-induced SOT pulse (Fig.~\ref{fig:exp+sim}a). The magnetization in the dot never fully switches, though, and images taken before and tens of nanoseconds after the pulse appear homogeneously magnetized. This observation matches our expectations from the simulation for this field regime, predicting only a transient coherent canting of spins in the dot. As the only notable difference, the simulated dynamics is faster than observed in the experiment, which we mainly attribute to the damping parameter in the simulations that was taken from literature and that we did not arbitrarily fine-tune to reproduce the experimental timescales (see Methods for simulation details). This apparently faster dynamics is consistently observed in all our simulations.

The type of dynamics changes dramatically when the applied field is reduced to a regime where switching becomes possible (Fig.~\ref{fig:exp+sim}b). In this case, the initialization current pulse in the pump--probe experiment nucleates the magnetic texture in the dot, which is then present at the beginning of all following cycles. Compared to the previous case, we therefore generally detect a substantially increased contrast as a result of the switched magnetization in the dot, in line with the simulations. Most strikingly, we observe SOT-driven deformation and elongation of the magnetization texture evolving from the dot into the surrounding non-modified magnetic film, culminating in shedding, where magnetic textures break-off and detach from the dot (Fig.~\ref{fig:exp+sim}b). Additional regimes presented in the Supplementary Fig.~2 include deformation and stripe-out dynamics. Across all cases, experimental observations are in good qualitative agreement with simulations.

\subsection{Magnetic instabilities and fluctuations}

\begin{figure*}
	\includegraphics[width=1.0\linewidth]{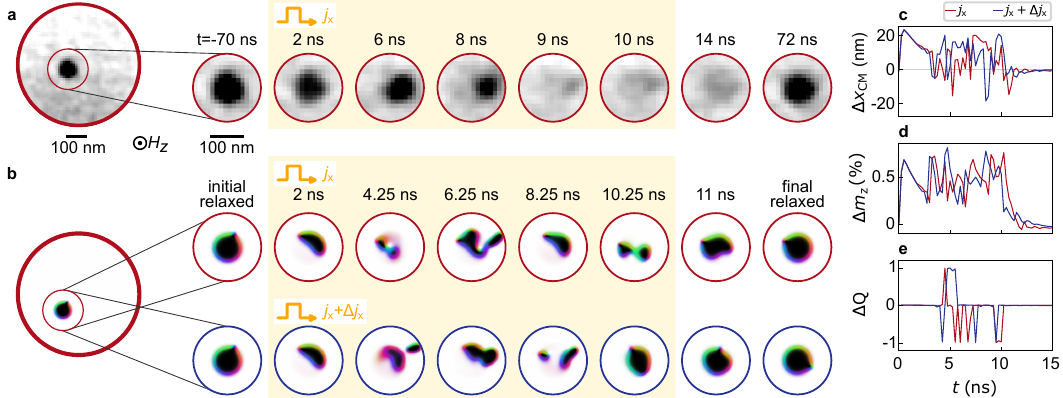} 
	\caption{\label{fig:chaos}\textbf{Transiently chaotic SOT dynamics.} \textbf{a} Time-resolved imaging results at $\mu_0 H_z = \SI{78}{mT}$, $j_x=j_0=\SI{4.3e11}{A \per m^2}$. The dynamics is restricted to the dot, and the averaged magnetization contrast reduces towards the end of the pulse. \textbf{b} Two simulation runs with identical initial state but slightly different current density with difference $\Delta j_x/j_x = 0.4\%$ and $j_x=2.5j_0$. \textbf{c--e} Analysis of the dynamic textures in the two simulation runs relative to the initial state. \textbf{c} The centre of mass displacement along the $x$-direction, \textbf{d} the magnetization difference (normalized to the \SI{600}{nm} field of view), and \textbf{e} change in skyrmion number.}
\end{figure*}

Despite the overall agreement between simulation and experiment, we observe one important difference: During the presence of the current-induced SOT pulse, the magnetization contrast in the experiment considerably reduces in the dot towards the end of the pulse. This does not only occur in the shedding regime (Fig.~\ref{fig:exp+sim}b), but is most notable at higher applied fields where the dynamics remain confined to the dot (Fig.~\ref{fig:chaos}a: washed out contrast for \SI{9}{ns} and \SI{10}{ns} delay). This finding is particularly surprising given that the skyrmion within the dot remains clearly visible both before and several nanoseconds after the current pulse. To better understand the experimentally observed contrast reduction evolving in the course of the current-induced SOT pulse (Fig.~\ref{fig:chaos}a), we revisited our simulations in the fluctuation prone regime (stripe-shaded region in Fig.~\ref{fig:phase_diagram}d), where the magnetization exhibits rapid transformations between transient textures that appear and vanish within a few nanoseconds---suggesting an inherently unstable and potentially chaotic character of the dynamics.

To test for chaos, we performed two micromagnetic simulations starting from identical initial states, varying only the current density slightly by \SI{0.4}{\percent}. The goal of this comparison is to probe the sensitivity of the system's evolution to infinitesimal perturbations, a classical hallmark of chaos. Although their early evolution during the current-induced SOT pulse is nearly indistinguishable for both runs (Fig.~\ref{fig:chaos}b), the magnetization patterns diverge dramatically for a delay of \SI{4.25}{ns} and later, producing qualitatively distinct, uncorrelated configurations despite the near-identical simulation conditions.

The divergence in transient magnetic states manifests itself in several observables. The center-of-mass position of the dynamic texture (Fig.~\ref{fig:chaos}c), the net magnetization contrast (Fig.~\ref{fig:chaos}d), and the skyrmion number (Fig.~\ref{fig:chaos}e) all fluctuate in a non-periodic and desynchronized manner between the two runs. Notably, the skyrmion number changes frequently during the pulse, pointing to a rich sequence of topological transitions and the formation of transient topologically skyrmion-like textures.

These micromagnetic simulations provide an explanation for our experimental observations: The x-ray images represent an average over millions of pump--probe cycles, each of which may follow a different dynamic pathway. If no pathway is predominant, the average image is washed out. While the magnetization dynamics in our simulations already show strong instabilities driven purely by the SOT, additional fluctuations such as thermal noise can further modify the trajectory taken in each cycle.
To assess the impact of such fluctuations, we further simulated seven sequential pulses with thermal noise (see Methods). Each relaxed state seeded the next cycle. These simulations reveal divergent magnetization pathways that almost consistently relax into a switched final state (except for one out of seven runs). Overlaying snapshots at selected timesteps (Supplementary Fig.~3) shows a pronounced washing-out of the magnetization, matching the reduced contrast observed during the current-induced SOT pulse in the experiment (Fig.~\ref{fig:chaos}b). Despite the variation across individual trajectories, recurring features---such as a rightward shift of the dark magnetic texture (Fig.~\ref{fig:chaos}c) and an increase in the out-of-plane magnetization component (Fig.~\ref{fig:chaos}d)---are consistently observed in both experiment and simulation. After the pulse, the magnetization contrast fully recovers, reflecting the return to a well-defined final state.

The SOT-driven shedding process (Fig.~\ref{fig:exp+sim}b) obeys the same dynamics: The stripe-out and break-off process is governed and driven by magnetic fluctuations as witnessed by the simulations and the faint contrast of the transient magnetic features. However, it must clearly contain a predominant ``core'' trajectory that consistently repeats in space and time, enabling consistent imaging of the break-off event, even though with reduced contrast. 

On the basis of our observation, we can rule out thermal demagnetization as the cause for the reduced contrast: Static heating from the injected electrical power would permanently erase the magnetic contrast. However, the contrast only diminishes after a few nanoseconds during the current pulse. Moreover, no contrast reduction is experimentally observed in stable regimes (e.g., canting or deformation), where the SOT drive induces minimal perturbations without entering a fluctuation regime. Therefore, we can also exclude transient heating that causes a purely thermal demagnetization as the main cause, supporting the view that transient contrast loss stems from fluctuations intrinsic to the SOT-driven regime.

\section*{Discussion}
A central feature of our experimental approach is the introduction of an anisotropy-engineered defect---the ``dot''---which acts as both the origin and return point for the SOT-induced magnetization dynamics. This design allows controlled and reproducible observations by spatially confining the dynamics and ensuring consistent initial and final magnetic states. As a result, the dot enables time-resolved, real-space access to so far inaccessible and unseen nano- to picosecond magnetization processes under well-defined conditions, offering a platform that can guide and host future studies on nanoscale magnetization dynamics.

We observe a rich variety of SOT-induced dynamics across a broad parameter space of current density and applied magnetic field in our ferromagnetic racetrack. At low current densities, the magnetization evolves deterministically, predictable and robust against perturbations, showing coherent canting and reversible deformations of magnetic textures. However, beyond a critical SOT threshold, the character of the dynamics changes abruptly to a transient regime dominated by rapid, disordered magnetization fluctuations. These include stripe-out and skyrmion shedding which we observed experimentally for the first time, and are characterized by topological transformations and divergent trajectories. These instabilities are intrinsic to the SOT-driven dynamics, and emerge in simulations even without thermal noise. The transient behavior becomes highly sensitive to small perturbations: even small variations in current or thermal fluctuations lead to divergent trajectories during the current-induced SOT pulse. The observed features---non-linear evolution, topological fluctuations, and sensitivity to initial conditions---are consistent with \textit{transient chaos}: a dynamical regime that is locally unpredictable, yet globally bounded in time and space.

Strikingly, the onset of fluctuation-driven behavior arises within a narrow window of current and magnetic field, suggesting the system operates near a dynamical critical point. In such a critical regime, complex behavior emerges naturally from deterministic dynamics, as expected near phase transitions in non-equilibrium systems~\cite{riego2018towards,Quintana2023Experimental}. Accordingly, the observed transition can be interpreted as a dynamic phase transition---from a dynamically ordered state to a transiently unstable, fluctuating state.

Crucially, the transient disorder observed in both experiment and simulation does not affect the final magnetization state, which remains reproducible across repetitions. Despite the intermediate instability, the magnetization contrast before and after the pulse is preserved---explaining why the pump--probe imaging remains viable. 
While averaging over many cycles suppresses transient single-cycle features, reproducible signatures such as directional motion and net magnetization increase persist, showing strong agreement between experiment and simulation. Thus, consistent final states can emerge from diverse and unpredictable transient pathways.

Note that a rigorous identification of \textit{chaos} based on Lyapunov exponents is not feasible in the present system. Such an analysis requires trajectory-resolved data over timescales significantly exceeding the Lyapunov time in order to ensure convergence of finite-time estimates. Given the nano- to picosecond intrinsic dynamics and the observed decorrelation times on the order of several nanoseconds, this would correspond to hundreds to thousands of characteristic cycles. These requirements are incompatible with both the experimentally accessible stroboscopic averaging and the computational cost of large-scale micromagnetic simulations. We therefore restrict our terminology to ``chaos-like'' dynamics, reflecting the observed sensitivity to perturbations and irregular, bounded behavior without claiming a formal mathematical proof of chaos.

Our findings challenge the view of SOTs as purely deterministic drivers and suggest that non-linear instabilities are a generic feature of driven magnetic systems in this regime. The current densities driving SOT in our experiment are frequently used in skyrmion experiments~\cite{buttner2017field, litzius2017skyrmion, woo2017spin, woo2018deterministic, finizio2019deterministic, juge2021helium, kern2022deterministic} and well below the destruction threshold of the device by Joule heating, even at MHz repetition rates (Supplementary Fig.~1d). That chaos-like dynamics arises under such moderate conditions underscores its immediate relevance for time-resolved studies and spintronics applications. 

\section*{Conclusion}
Our findings situate skyrmion dynamics within the wider context of SOT-driven magnetization phenomena, including soliton emission~\cite{mohseni2013spin,bonetti2015direct}, spin-wave generation~\cite{fulara2019spin}, and auto-oscillations~\cite{demidov2020spin}. Moreover, our findings underscore the rich and often hidden complexity underlying magnetically ordered systems~\cite{Fanciulli2025Magnetic}. We highlight the dual nature of SOTs as drivers of both deterministic and complex non-linear behavior. While the pursuit of precise control in magnetic systems remains central, particularly for applications, our work suggests that SOT-driven transient chaos could be harnessed as a powerful mechanism for enabling topological transitions and texture generation via skyrmion shedding, effectively bridging spin-wave and skyrmion dynamics on short timescales, without the need for an assisting in-plane field or edge-mediated effects~\cite{baumgartner2017spatially}. Our work invites further exploration of the role of instabilities regarding both the physics and the possible functionalities of driven spin textures, leveraging chaos-based~\cite{ditto2008chaos} and probabilistic computing architectures~\cite{daniels2020energy, talatchian2021mutual}.

\section{Methods}\label{sec:methods} 
\subsection{Sample fabrication}
Ferromagnetic multilayers with a nominal composition of Ta(\SI{3}{nm})/Pt(\SI{4}{nm})/ \newline[Pt(\SI{2.5}{nm})/Co$_{60}$Fe$_{25}$B$_{15}$(\SI{0.7}{nm})/MgO(\SI{1.4}{nm})]$_{15}$/Pt(\SI{2}{nm}) were deposited on \SI{500}{nm} thick Si$_{3}$N$_4$-membranes via argon-ion-assisted DC and RF magnetron sputtering. The magnetization curve, as shown in Supplementary Fig.~1a, was recorded using magneto-optical Kerr microscopy. The domain width at remanence was determined via magnetic force microscopy to be $\approx$\SI{120}{nm}. In a magnetic film with the same nominal composition from a different fabrication run, we obtained a saturation magnetization of $M_S=1.28\times 10^{6}$ A/m  and an anisotropy constant $K_\mathrm{u}=$\SI{1.55}{MJ/m}$^3$ via SQUID magnetometry. To allow for current injection into the device, the magnetic film was structured into a magnetic racetrack geometry with a narrowed region of \SI{3}{\micro m} $\times$ \SI{1}{\micro m}, using a combination of UV lithography and focused Ga$^+$-ion-beam milling (FEI Helios Nanolab \num{600}) (see Supplementary Fig.~1b). In addition, employing UV lithography and electron-beam evaporation, gold contacts were deposited to the sides of the racetrack for wire bonding.

For x-ray holography, the field of view is fixed to a circular region of \SI{600}{nm} diameter defined by an otherwise x-ray opaque \SI{1}{\micro m} thick gold mask layer on the back side of the SiN membrane. The monolithic integrated sample design \cite{eisebitt2004lensless} allows for drift-free operation \cite{buttner2015dynamics}. Six pinholes of varying diameter between \num{20} and \SI{110}{nm} were placed around the circular aperture to provide holographic reference beams.

Next, the magnetic film was locally irradiated using a focused helium ion beam, similar to the work presented in Ref.~\citenum{kern2022deterministic}. Employing focused ion beams has the great advantage that preferred nucleation sites can be patterned with a high degree of control in terms of their shape and position as well as their anisotropy reduction. We prepared a circular area of \SI{50}{nm} radius with an ion dose of $125$ ions/nm$^2$, using the ZEISS Orion NanoFab He$^+$-FIB and the FIB-o-mat software package \cite{deinhart2021patterning}. Using a custom-developed sample aligning tool, FIB patterns are written with a placement accuracy of at least \SI{200}{nm} by employing prerecorded optical and scanning electron micrographs for sample navigation, avoiding any inadvertent ion irradiation. The irradiated areas are composed of filled circular shapes which are rasterized in concentric circles with a dwell time of \SI{0.1}{\micro s} and a pitch of \SI{1}{nm} in $x$- and $y$-directions. For the patterning process, we used an acceleration voltage of \SI{30}{kV}, a \SI{20}{\micro m} wide aperture and ion currents between \SI{2.5}{pA} and \SI{4}{pA}. Note here that we denote disks of \SI{50}{nm} radius as ``dots'' throughout the main text. 

The multilayer is entirely field-polarized at an applied field above $\mu_0 H_\mathrm{S} = \SI{110}{mT}$ (see magnetization curve in Supplementary Fig.~1). Lowering the applied field below $\mu_0 H_\mathrm{N} = \SI{78}{mT}$, the magnetization in the He$^+$-irradiated region switches spontaneously before the rest of the film eventually breaks into magnetic domains. Even above $\mu_0H_\mathrm{N}$, a single spin--orbit torque pulse generated from electric current can still induce local magnetization switching in this region, resulting in the formation of a skyrmion \cite{kern2022deterministic}. At $\mu_0 H_z=\SI{78}{mT}$, we find the threshold current density to nucleate a skyrmion in the ion-irradiated region to be $j_0 \equiv j_x=\SI{4.3e11}{A \per m^2}$. Throughout this work, we express the current density with respect to this reference current $j_0$.

\subsection{X-ray imaging}
The magnetic imaging was carried out using resonant coherent x-ray imaging based on a holography in transmission geometry \cite{eisebitt2004lensless}. By tuning the x-ray wavelength to the Co L$_3$ resonance (at \SI{1.59}{nm}), the x-ray magnetic circular dichroism provides contrast to the out-of-plane component of the magnetization ($m_z$). We performed the experiment at the MAX-P04 holography endstation at beamline P04 of the synchrotron-radiation facility PETRA III, Hamburg. Combining the holographic image retrieval with refinements via phase-retrieval algorithms, we achieve almost diffraction-limited spatial resolution of \SI{18}{nm} \cite{battistelli2023coherent} and a temporal resolution limited by the x-ray pulse duration of \SI{100}{ps}. During the imaging experiments, a variable out-of-plane magnetic field ($H_{\mathrm{z}}$) was applied by a dipole electromagnet.

In order to perform a time-resolved pump--probe imaging experiment, we synchronize a nanosecond-pulsed current excitation as the pump pulse to the soft-x-ray probe pulse at a repetition rate of \SI{5.2}{MHz}, leading to a cycle period of \SI{192}{ns}. Figure~\ref{fig:setup}b schematically shows the experimental setup with the magnetic sample in the center. A programmable pulse generator (HP 8131A) injects current pulses into the racetrack device. This pulse generator provides up to \SI{500}{MHz} repetition rates, \SI{500}{ps}--\SI{99.9}{ms} pulse widths with \SI{10}{ps} timing resolution, $<$\SI{200}{ps} rise/fall times, and up to \num{5} Vpp output into \SI{50}{\ohm}, making it well suited for fast and precise current-pulse experiments. The arrival of the x-ray pulses at the sample (time-zero) is determined with an avalanche photodiode mounted at the sample position. A back-illuminated CCD ($2048\times 2048$ pixels) recorded the x-ray holograms with an accumulated exposure time of \SI{9}{s} per delay step, which corresponds to an averaging over \num{4.7e7} pump--probe cycles in a single image. 

We additionally carried out finite-element simulations using a commercial-grade backward differentiation formula solver (COMSOL multiphysics package, COMSOL AB, Stockholm) to estimate the expected heat load on the sample induced by the pulsed current drive with a period of \SI{192}{ns}. We combined the simulation of the electric potential with a subsequent heat transfer across the interfaces via an electromagnetic coupling, based on the time-dependent heat equation in a 2D model of the sample. We simulate the multilayer as one layer, and therefore use the current density normalized to the full stack, $j_x=\SI{2.56e11}{A \per m^2}$. Based on these finite-element simulations run for \SI{100}{\micro s}, we estimate the Joule heating of our samples during the pump-probe imaging for the current density $j_0$ (see Supplementary Fig.~1d). While the sample transiently reaches a temperature of \SI{22}{K} above room temperature, we obtain a persistent residual heating of \SI{9}{K} at our repetition rate. We do not expect such low residual heat to significantly alter our experimental observations.

\subsection{Micromagnetic simulations}
Micromagnetic simulations were performed with the MicroMagnum framework and custom extensions for the DMI and SOT. All parameters were scaled according to the effective medium approach \cite{woo2016observation} to obtain the dynamics of the 15-layer stack in a single simulated layer. The simulated area consisted of a cartesian grid with $1000 \times 1000 \times1$ cells in $xyz$-direction. Each cell has a lateral dimension of $1\,\mathrm{nm}\times1\,\mathrm{nm}\times78\,\mathrm{nm}$, where the extension in $z$-direction corresponds to the thickness of the entire experimental multilayer stack. 

We used the following parameters: Gilbert damping $\alpha = 0.1$, saturation magnetization $M_\mathrm{S} = \SI{1.28e6}{A/m}$, exchange stiffness $A = \SI{10}{pJ/m}$, DMI constant $D = \SI{-1.5}{mJ/m^2}$, and effective anisotropy field $H_K = \SI{0.81}{T}$. The irradiated area of lower anisotropy was modeled by a circular area of $\SI{100}{nm}$ diameter in the center of the simulated square. 
Within the area, the spin Hall angle $\theta_\mathrm{H} = 0.15$ was used together with a reduced uniaxial anisotropy of $0.2 K_\mathrm{u}$. The outside was modeled with lower spin-torque efficiency of $\theta_\mathrm{H} = 0.1$ and the full $K_\mathrm{u}$. 
Helium-ion irradiation is known to induce interfacial intermixing and atomic displacements. Previous work~\cite{chappert1998planar,sud2021tailoring} shows that the anisotropy constant $K_{\mathrm{u}}$ can decrease substantially (up to $\approx 60\%$), whereas the Dzyaloshinskii--Moriya interaction remains largely unchanged. Previous work also demonstrated that helium ion irradiation in these [Pt/CoFeB/MgO] thin films enables a wide dose window for tuning magnetic properties while preserving out-of-plane magnetized textures~\cite{kern2022deterministic}. In addition helium ion irradiation increases the spin Hall angle, thereby enhancing spin-orbit torque efficiency and reducing switching currents~\cite{dunne2020helium}. Accordingly, we also accounted for this by increasing the spin Hall angle in the anisotropy-engineered dot by 50\%. Since nanoscale measurements of irradiation-induced changes are not feasible with techniques such as SQUID magnetometry, we rely on established full-film studies.

Note that we did not attempt to arbitrarily adjust the material parameters to further match the simulation with the experiment and relied on measurements and literature values for the experimental system used. The only deviation from the experiment we had to include is a factor of \num{2.5} higher current densities in the simulations to achieve the same response as in the experiment. This might be caused by a different spin-torque efficiency in our (irradiated) material compared to literature values.

The simulations encompassed exchange, anisotropy, stray field, DMI, and external field modules as well as a SOT of Slonczewski type. We neglect the field-like SOT and focus on the damping-like contribution, which provides the dominant driving force for skyrmion motion. Within the Thiele formalism, only the damping-like SOT generates an effective force, whereas the field-like torque acts as an in-plane field and does not couple to the translational mode~\cite{manchon2019current}. It therefore primarily modifies the spin texture without contributing to sustained dynamics in the present context. The initial state was chosen to be empty and relaxed before applying a 10-ns current pulse. After the first current pulse, the system was relaxed to find the ground state before applying the second consecutive pulse of \SI{10}{ns}. This second pulse constitutes an individual cycle of the pump--probe scheme. Finally, a third relaxation was performed to find the new relaxed state. This was done to account for the experimental pump--probe method that chains together many cycles during a single recorded frame, effectively averaging them, and thus putting very little weight on the first initial, nucleating cycle. We, thus, only refer to the second simulation cycle when comparing simulations to the experimental results. Over a large range in our phase diagram in Fig.~\ref{fig:phase_diagram}d, we find relaxed states mostly consistent with the initial state of the second spin-orbit torque pulse. However, we observed two exceptions at the boundary between the regime of magnetic fluctuations and the canting regime. In these cases, the system relaxed to a non-switched state when initially started in a switched state. We avoided this region in the experiment.

Simulating an individual pump--probe cycle, we can investigate a single signature of the magnetic response to the current pump pulse, which complements our experimental findings. In addition, we provide the simulation of seven subsequent pulses in Supplementary Fig.~3: Adding the stochasticity of a temperature ``field'' \cite{Leliaert_SimTemperature} during the first \num{10} calculation steps, we prepare infinitesimally changed initial states at the onset of each SOT pulse. Note that introducing stochasticity in the simulations is not feasible to the extent our experiment can provide it. Increasing the number of simulated pump--probe cycles is computationally very expensive and does not further advance our insights since it cannot account for slight temperature variations or magnetic inhomogeneities always present in the material.

\section*{Supporting Information}
Supplementary Material includes Supplementary Figures of the magnetization curve of the magnetic film, the experimentally applied current pulse shape, the calculated current-induced temperature evolution, a time series for deformation dynamics, for stripe-out dynamics and chaotic fluctuations; Movies of the magnetization dynamics as discussed in the main paper Fig.~\ref{fig:phase_diagram} and Fig.~\ref{fig:exp+sim}.

\section*{Acknowledgements}
The measurements presented in this work were carried out at PETRA III (DESY). We acknowledge DESY (Hamburg, Germany), a member of the Helmholtz Association HGF, for the provision of experimental facilities at PETRA III, beamline P04. The ion-beam patterning was performed in the Corelab Correlative Microscopy and Spectroscopy at Helmholtz-Zentrum Berlin (HZB). Financial support from the Leibniz Association via Grant No. K162/2018 (OptiSPIN) and K720/2025 (X-MAG) and the Helmholtz Young Investigator Group Program is acknowledged. In addition, we would like to acknowledge the support from the EU COST Action CA 19140 (FIT4NANO). L.-M.K. acknowledged financial support from the Deutsche Forschungsgemeinschaft (DFG, German Research Foundation) - project number 546268067.

\section*{Author contributions}
L.-M.K. and B.P. conceived and designed the experiment. L.-M.K., D.E., M.S., C.M.G., V.D., and K.H. prepared and characterized the samples. L.-M.K., V.D., M.S., C.K., R.B., D.M., and M.-J.H. performed the experiment. L.-M.K., C.K., K.G., and R.B. reconstructed the holographic images. K.L. performed the micromagnetic simulations. L.-M.K., K.L., F.B., S.E., and B.P. interpreted the results. L.-M.K. and K.L. drafted the manuscript and prepared the figures.  B.P., F.B., and S.E. supervised the project. All authors discussed the results and commented on the manuscript.

\section*{Competing financial interests}
The authors declare no competing financial interests.

\section*{Data availability}
The data from experiment and simulation that support the findings and figures within this paper are available from the corresponding author on reasonable request.

%

\end{document}